\documentclass{article}
\usepackage{spconf,amsmath,graphicx}
\usepackage{graphicx,algorithm,algorithmicx,bm,amsmath,amssymb,rotating,graphicx,cite,amsthm,calc,color,epstopdf,xcolor,mathrsfs,dsfont}
\usepackage{lettrine}
\usepackage{cite}
 \usepackage{amsfonts}
\usepackage{longtable}
\usepackage{float}
\usepackage{subcaption}
\usepackage{flushend}
\usepackage{multirow}
\definecolor{ICIP}{rgb}{0.21,0.49,0.74}
\usepackage[breaklinks,colorlinks,citecolor=ICIP]{hyperref}
\usepackage{cleveref}
\usepackage{booktabs}

\begin{document}

\title{Advancements in Medical Image Classification through Fine-Tuning Natural Domain Foundation Models }

\name{Mobina Mansoori$^\dag$, Sajjad Shahabodini$^\dag$, Farnoush Bayatmakou$^\dag$, Jamshid Abouei $^{\dag\dag}$,\vspace{-.14in}}
\address{\textit{Konstantinos N. Plataniotis$^\ddag$, and Arash Mohammadi$^\dag$}\\\\
$~^\dag$ Intelligent Signal \& Information Processing (I-SIP) Lab, Concordia University, Canada \\$~^\ddag$ Department of Electrical and Computer Engineering, University of Toronto \\$~^{\dag\dag}$ Department of Electrical Engineering, Yazd University, Iran}

\ninept
\maketitle

\begin{abstract}
Using massive datasets, foundation models are large-scale, pre-trained models that perform a wide range of tasks. These models have shown consistently improved results with the introduction of new methods. It is crucial to analyze how these trends impact the medical field and determine whether these advancements can drive meaningful change. This study investigates the application of recent state-of-the-art foundation models—DINOv2, MAE, VMamba, CoCa, SAM2, and AIMv2—for medical image classification. We explore their effectiveness on datasets including CBIS-DDSM for mammography, ISIC2019 for skin lesions, APTOS2019 for diabetic retinopathy, and CHEXPERT for chest radiographs. By fine-tuning these models and evaluating their configurations, we aim to understand the potential of these advancements in medical image classification. The results indicate that these advanced models significantly enhance classification outcomes, demonstrating robust performance despite limited labeled data. Based on our results, AIMv2, DINOv2, and SAM2 models outperformed others, demonstrating that progress in natural domain training has positively impacted the medical domain and improved classification outcomes. Our code is publicly available at \url{https://github.com/sajjad-sh33/Medical-Transfer-Learning}.
\end{abstract}

\begin{keywords}
Transfer Learning, Medical Image Classification, Multi-Modal Models, Foundation Models
\end{keywords}

\section{Introduction}
Foundation models are large-scale, pre-trained models designed to perform a wide range of tasks by learning from massive datasets. In recent years, foundation models have gained significant interest in the research community, becoming a cornerstone in various fields of artificial intelligence \cite{bommasani2021opportunities}. Their development began with groundbreaking works in natural language processing (NLP), such as GPT (Generative Pre-trained Transformer)\cite{radford2019language} and BERT (Bidirectional Encoder Representations from Transformers)\cite{kenton2019bert}, which revolutionized text understanding and generation. These models, including the famous ChatGPT, demonstrate the efficacy of pre-training vast amounts of data and fine-tuning for specific tasks. Their remarkable capability to generalize across diverse tasks has led to an increase in their adoption, with a growing number of papers leveraging these models for their specific applications.

Building on the success of NLP foundation models, researchers extended these concepts to computer vision, resulting in significant advancements over the past few years. Models like CLIP (Contrastive Language-Image Pre-Training)\cite{radford2021learning}, MAE (Masked Autoencoders)\cite{he2022masked}, DINOv2\cite{oquab2023dinov2}, VMamba\cite{liu2024vmamba}, SAM2 (Segment Anything 2)\cite{ravi2024sam}, AIMv2\cite{fini2024multimodal}, and CoCa (Cooperative Convolutional Attention)\cite{yu2022coca} have set brand-new benchmarks for various visual tasks. These computer vision foundation models exhibit robust performance, even in scenarios involving limited labeled data, by leveraging pre-trained knowledge.

The medical imaging field often encounters difficulties due to a lack of labeled data, complicating the process of training models from scratch. Transfer learning offers an effective solution by fine-tuning pre-trained models for specific tasks. These foundation models, with extensive pre-training on large-scale datasets, can transfer learned representations to new tasks, proving highly advantageous for medical image classification where annotated data is scarce\cite{matsoukas2023pretrained,matsoukas2022makes}. However, domain shifts between the pre-training data and the target medical tasks pose a significant hurdle, necessitating the evaluation of these models' effectiveness in this specific field. While attempts have been made to adapt previous foundation models to the medical domain\cite{huix2024natural,shi2023generalist,wang2022medclip,mansoori2024polyp,yi2023towards,mansoori2024self,SAM2-MIS}, this study focuses on recent state-of-the-art models in natural domain image classification to demonstrate how advancements in this area can impact medical image classification outcomes and shows how trends in naturally trained models affect the medical domain.

In this study, we aim to bridge the existing gap by conducting a thorough evaluation and analysis of advanced foundation models in the context of medical image classification. Our goal is to highlight their transferability and effectiveness in the medical domain. Our key findings indicate that advancements in foundation models continue to yield significant results in the medical domain. We found that almost newer models achieved better outcomes compared to older ones while reducing the number of parameters. This demonstrates a direct correlation between progress in both the medical and natural domains. Additionally, while foundation models perform well with frozen features for specific tasks, fine-tuning is essential for optimal performance in medical applications.

\section{METHODOLOGY}
Our methodology explores the potential of fine-tuning advanced foundation models such as CoCa, VMamba, AIMv2, MAE, SAM2, DINOv2, and CLIP to improve the performance and reliability of medical image classification tools. By fine-tuning these models on specific medical datasets, we aim to uncover their effectiveness in distinguishing various medical conditions, focusing on tasks like breast cancer diagnosis, diabetic retinopathy detection, skin lesion classification, and automated chest X-ray interpretation. As a follow-up to \cite{huix2024natural}, different configurations, including freezing versus unfreezing the models and utilizing either linear heads or multi-layer attention heads, are compared to evaluate their performance.
\subsection{Foundation Models}
In this study, we evaluate the performance of seven different foundation models on medical image datasets when fine-tuned. We use models that are trained with various pre-training approaches, including label-supervised, text-to-image supervised, mask-supervised, and self-supervised techniques. We focus on recently published models, while also including DINOv2 and CLIP, which have previously shown great potential compared to former state-of-the-art models. A brief overview of these models follows:

•	\textbf{CLIP}\cite{radford2021learning}: CLIP is a neural network trained on a vast dataset of 400 million (image, text) pairs collected from the internet. It employs a Vision Transformer (ViT) \cite{dosovitskiy2020image} to extract visual features and a Transformer language model \cite{radford2019language} to extract text features. Both features are projected into a shared latent space, where the dot product between the projected image and text features is used as a similarity score. CLIP can perform a wide range of classification benchmarks without optimizing them directly. We conducted our experiments using the ViT-B version of this model. 

•	\textbf{DINOv2}\cite{oquab2023dinov2}: DINOv2 is an advanced self-supervised learning method applied to ViTs. It has been trained on a curated dataset of 142 million images without using any labels or annotations. The model produces robust visual features that can be directly employed with simple classifiers for various computer vision tasks. DINOv2 achieves impressive performance on benchmarks like ImageNet-1k\cite{matsoukas2022makes} and MSCOCO\cite{lin2014microsoft} object detection, surpassing its previous state-of-the-art models. We use the ViT-B and ViT-L models as the backbone for feature extraction. 

\begin{table*}[h!]
\caption{Experimental results on the CBIS-DDSM\cite{lee2017curated} dataset. We use the ROC-AUC metric for the evaluation.}
\centering
\fontsize{10pt}{8pt}\selectfont
\renewcommand{\arraystretch}{1}
\setlength{\tabcolsep}{24pt}
\begin{tabular}{c| c| c| c| c| c c c}
    \toprule
    \multirow{2}{*}{\textbf{Methods}} & \multirow{2}{*}{\textbf{Backbones}} & \multicolumn{2}{c|}{\textbf{Frozen}} & \multicolumn{2}{c}{\textbf{Unfrozen}} \\
    \cmidrule{3-4} \cmidrule{5-6}
    & & \textit{Linear} & \textit{Attention} & \textit{Linear} & \textit{Attention} \\
        \midrule
{\textbf{CLIP}\cite{radford2021learning}} & \textit{ViT-B}&  0.897& 0.948  &  0.945& 0.955  \\
    \midrule
\multirow{2}{*}{\textbf{DINOv2}\cite{oquab2023dinov2}} & \textit{ViT-B}&0.905 & 0.945  &0.966&\textbf{0.966}  \\
& \textit{ViT-L}&0.914 & \textbf{0.954}  &0.965&\textbf{0.966}  \\
    \midrule
\multirow{3}{*}{\textbf{MAE}\cite{he2022masked}} & \textit{ViT-B}&0.870 &0.937  &0.943&0.942  \\
& \textit{ViT-L}&0.872 &0.931  &0.945 &0.944  \\
& \textit{ViT-H} &0.865&0.939 &0.941 &0.941  \\
\midrule
\multirow{2}{*}{\textbf{VMamba}\cite{liu2024vmamba}}& \textit{VMamba-S} &0.869 &0.931 &0.938 &0.935  \\
& \textit{VMamba-B} &0.871 &0.935 &0.941 &0.940 \\
\midrule
\multirow{2}{*}{\textbf{CoCa}\cite{yu2022coca} }& \textit{ViT-B} &0.863 &0.929 &0.939 &0.938  \\
& \textit{ViT-L} &0.864 &0.930 &0.942 &0.941  \\
\midrule
\multirow{2}{*}{\textbf{SAM2}\cite{ravi2024sam} }& \textit{Hiera-B} &0.888&0.932 &0.952 &0.941  \\
& \textit{Hiera-L} &0.889 &0.937 &0.954 &0.943  \\
\midrule
\multirow{2}{*}{\textbf{AIMv2}\cite{fini2024multimodal}} & \textit{ViT-L}&0.898 &0.948  &\textbf{0.968} &\textbf{0.966}  \\
& \textit{ViT-H} & \textbf{0.915}&\textbf{0.954}&0.966 &0.965  \\
\midrule
\end{tabular}
\label{ddsm}
\end{table*}

\begin{table*}[h!]
\caption{Experimental results on the  APTOS2019\cite{karthik2019aptos} dataset. We use the Quadratic Cohen Kappa metric for the evaluation.}
\centering
\fontsize{10pt}{8pt}\selectfont
\renewcommand{\arraystretch}{1}
\setlength{\tabcolsep}{24pt}
\begin{tabular}{c| c| c| c| c| c c c}
    \toprule
    \multirow{2}{*}{\textbf{Methods}} & \multirow{2}{*}{\textbf{Backbones}} & \multicolumn{2}{c|}{\textbf{Frozen}} & \multicolumn{2}{c}{\textbf{Unfrozen}} \\
    \cmidrule{3-4} \cmidrule{5-6}
    & & \textit{Linear} & \textit{Attention} & \textit{Linear} & \textit{Attention} \\
    \midrule
    {\textbf{CLIP}\cite{radford2021learning}} & \textit{ViT-B}&   0.857&  0.903  &   0.903& \textbf{0.907}  \\
    \midrule
\multirow{2}{*}{\textbf{DINOv2}\cite{oquab2023dinov2}} & \textit{ViT-B}& 0.881 &   0.901  & 0.909& 0.904  \\
 &\textit{ViT-L}& \textbf{0.889}&   0.905  & 0.913& 0.905  \\
    \midrule
\multirow{3}{*}{\textbf{MAE}\cite{he2022masked}} & \textit{ViT-B}&0.882 &0.915  &0.903 &0.902  \\
& \textit{ViT-L}&0.884 &0.906  &0.904 &0.903  \\
& \textit{ViT-H}&0.881 &0.901  &0.901 &0.901  \\
\midrule
\multirow{2}{*}{\textbf{VMamba}\cite{liu2024vmamba}}  & \textit{VMamba-S} &0.863 &0.885&0.872 &0.886  \\
& \textit{VMamba-B} &0.866 &0.887 &0.879 &0.887  \\
\midrule
\multirow{2}{*}{\textbf{CoCa}\cite{yu2022coca} }& \textit{ViT-B} &0.879 &0.899 &0.901 &0.900 \\
& \textit{ViT-L} &0.882 &0.903&0.902 &0.902 \\
\midrule
\multirow{2}{*}{\textbf{SAM2}\cite{ravi2024sam} }& \textit{Hiera-B} &0.878 &0.895 &0.905 &0.903  \\
& \textit{Hiera-L} &0.881 &0.899 &0.908 &0.904  \\
\midrule
\multirow{2}{*}{\textbf{AIMv2}\cite{fini2024multimodal} }& \textit{ViT-L}&0.888 &\textbf{0.916}  &\textbf{0.915} &\textbf{0.907}  \\
& \textit{ViT-H} &0.885&0.914 & 0.909&0.906  \\
\midrule
\end{tabular}
\label{aptos}
\end{table*}

•	\textbf{MAE}\cite{he2022masked}: MAE leverages an encoder-decoder framework tailored to representation learning tasks. A total of 130 million images were used to train this model, which includes publicly available datasets such as ImageNet\cite{deng2009imagenet} and MSCOCO. During training, random patches of the images are masked, and the model aims to reconstruct the original image by understanding the visual features and patterns within the data. For this model, our implementation uses three main backbones, including ViT-B, ViT-L, and ViT-H. 

•	\textbf{VMamba}\cite{liu2024vmamba}: Visual State Space Model is a cutting-edge neural network designed to analyze complex visual data. It adapts Mamba \cite{gu2023mamba}, a state-space language model, to a vision backbone with linear time complexity. VMamba excels at tasks such as object detection, segmentation, and image classification. Training on diverse datasets makes it a powerful tool for visual data analysis. Our experimental implementation employs both VMamba-S and VMamba-B versions.

•	\textbf{CoCa}\cite{yu2022coca}: Contrastive Captioner is the current state-of-the-art Image-Text Foundation Model designed to bridge the gap between visual and textual information. CoCa model leverages contrastive learning techniques to align image and text representations, facilitating tasks like image captioning, visual question answering, and multimodal understanding. These models have been trained on vast datasets of image-text pairs, enabling them to generate accurate and contextually relevant captions for a wide range of visual inputs. Our implementation integrates the ViT-B and ViT-L architecture backbones. 

•	\textbf{SAM2} \cite{ravi2024sam}: Segment Anything 2 is an advanced foundation model designed for prompt segmentation of both images and videos. Built on a simple transformer architecture with streaming memory, SAM2 supports real-time video segmentation and has been trained on the largest available segmentation dataset. It demonstrates strong performance across various tasks and visual domains, making it a significant milestone in visual segmentation technology.

•	\textbf{AIMv2}\cite{fini2024multimodal}: The AIMv2 model adopts a multimodal autoregressive pre-training technique, combining a substantial vision encoder with a multimodal decoder that outputs both raw image patches and text tokens. This model is trained on an extensive collection of 142 million image and text pairs, drawn from well-known datasets. AIMv2 excels across various downstream tasks such as localization, grounding, and classification.

\begin{table*}[h!]
\caption{Experimental results on the  ISIC2019\cite{codella2018skin} dataset. We use the recall metric for the evaluation.}
\centering
\fontsize{10pt}{8pt}\selectfont
\renewcommand{\arraystretch}{1}
\setlength{\tabcolsep}{24pt}
\begin{tabular}{c| c| c| c| c| c c c}
    \toprule
    \multirow{2}{*}{\textbf{Methods}} & \multirow{2}{*}{\textbf{Backbones}} & \multicolumn{2}{c|}{\textbf{Frozen}} & \multicolumn{2}{c}{\textbf{Unfrozen}} \\
    \cmidrule{3-4} \cmidrule{5-6}
    & & \textit{Linear} & \textit{Attention} & \textit{Linear} & \textit{Attention} \\
    \midrule
    {\textbf{CLIP}\cite{radford2021learning} }& \textit{ViT-B}&   0.489& 0.748  &   0.818& 0.809  \\
    \midrule
 \multirow{2}{*}{\textbf{DINOv2}\cite{oquab2023dinov2}} & \textit{ViT-B}&0.569 &  0.790  &0.859&0.822  \\
 & \textit{ViT-L}&0.625&  0.791  &0.861&\textbf{0.854 }  \\
    \midrule
\multirow{3}{*}{\textbf{MAE}\cite{he2022masked}} & \textit{ViT-B}& 0.576&0.762  &0.844 &0.801 \\
& \textit{ViT-L}&0.584 &0.796  &0.846 &0.819  \\
& \textit{ViT-H} &0.623 &0.802 &0.847 &0.823  \\
\midrule
\multirow{2}{*}{\textbf{VMamba}\cite{liu2024vmamba}} & \textit{VMamba-S} &0.397 &0.736 &0.755 &0.759  \\
& \textit{VMamba-B} &0.401 &0.738 &0.763 &0.765  \\
\midrule
\multirow{2}{*}{\textbf{CoCa}\cite{yu2022coca}} & \textit{ViT-B} &0.574 &0.761 &0.837 &0.787  \\
& \textit{ViT-L} &0.581 &0.792 &0.839 &0.809  \\
\midrule
\multirow{2}{*}{\textbf{SAM2}\cite{ravi2024sam}} & \textit{Hiera-B} &0.447 &0.745 &0.845 &0.798  \\
& \textit{Hiera-L} &0.458 &0.747 &0.848 &0.801  \\
\midrule
\multirow{2}{*}{\textbf{AIMv2}\cite{fini2024multimodal}} & \textit{ViT-L}&0.583 &0.762  &0.860 &0.827  \\
& \textit{ViT-H} &\textbf{0.631}&\textbf{0.795} &\textbf{0.863}&\textbf{0.854 } \\
\midrule
\end{tabular}
\label{isic}
\end{table*}

\begin{table*}[h!]
\caption{Experimental results on the CHEXPERT\cite{irvin2019chexpert} dataset. We use the ROC-AUC metric for the evaluation.}
\centering
\fontsize{10pt}{8pt}\selectfont
\renewcommand{\arraystretch}{1}
\setlength{\tabcolsep}{24pt}
\begin{tabular}{c| c| c| c| c| c c c}
    \toprule
    \multirow{2}{*}{\textbf{Methods}} & \multirow{2}{*}{\textbf{Backbones}} & \multicolumn{2}{c|}{\textbf{Frozen}} & \multicolumn{2}{c}{\textbf{Unfrozen}} \\
    \cmidrule{3-4} \cmidrule{5-6}
    & & \textit{Linear} & \textit{Attention} & \textit{Linear} & \textit{Attention} \\
    \midrule
    \multirow{1}{*}{\textbf{CLIP}\cite{radford2021learning} }& \textit{ViT-B}& 0.702&   0.790  &   0.806& 0.805  \\
    \midrule
\multirow{2}{*}{\textbf{DINOv2}\cite{oquab2023dinov2}} & \textit{ViT-B}&  0.722 &   0.798 &  0.812&  0.812  \\
 & \textit{ViT-L}&  \textbf{0.731} &   0.801 &  0.814&  0.813 \\
    \midrule
\multirow{3}{*}{\textbf{MAE}\cite{he2022masked} }& \textit{ViT-B} &0.713 &0.793 &0.802 &0.801  \\
& \textit{ViT-L} &0.715 &0.801 &0.805 &0.805  \\
& \textit{ViT-H} &0.725 &0.802 &0.807 &0.809 \\
\midrule
\multirow{2}{*}{\textbf{VMamba}\cite{liu2024vmamba}} & \textit{VMamba-S} &0.715 &0.773 &0.777 &0.775  \\
& \textit{VMamba-B} &0.717 &0.778 &0.779 &0.778  \\
\midrule
\multirow{2}{*}{\textbf{CoCa}\cite{yu2022coca}} & \textit{ViT-B} &0.708 &0.781 &0.801 &0.800  \\
& \textit{ViT-L} &0.711 &0.792 &0.805 &0.804  \\
\midrule
\multirow{2}{*}{\textbf{SAM2}\cite{ravi2024sam} }& \textit{Hiera-B} &0.728 &0.798 &0.798 &0.794  \\
& \textit{Hiera-L} &0.729 &0.793 &0.801 &0.793  \\
\midrule
\multirow{2}{*}{\textbf{AIMv2}\cite{fini2024multimodal} }& \textit{ViT-L} &0.724 &0.803 &0.813 &0.814  \\
& \textit{ViT-H} &\textbf{0.731}  &\textbf{0.807}  &\textbf{0.819} & \textbf{0.815 } \\
\midrule
\end{tabular}
\label{chexpert}
\end{table*}

\begin{figure*}[h!]
\centering
\includegraphics[width=\textwidth]{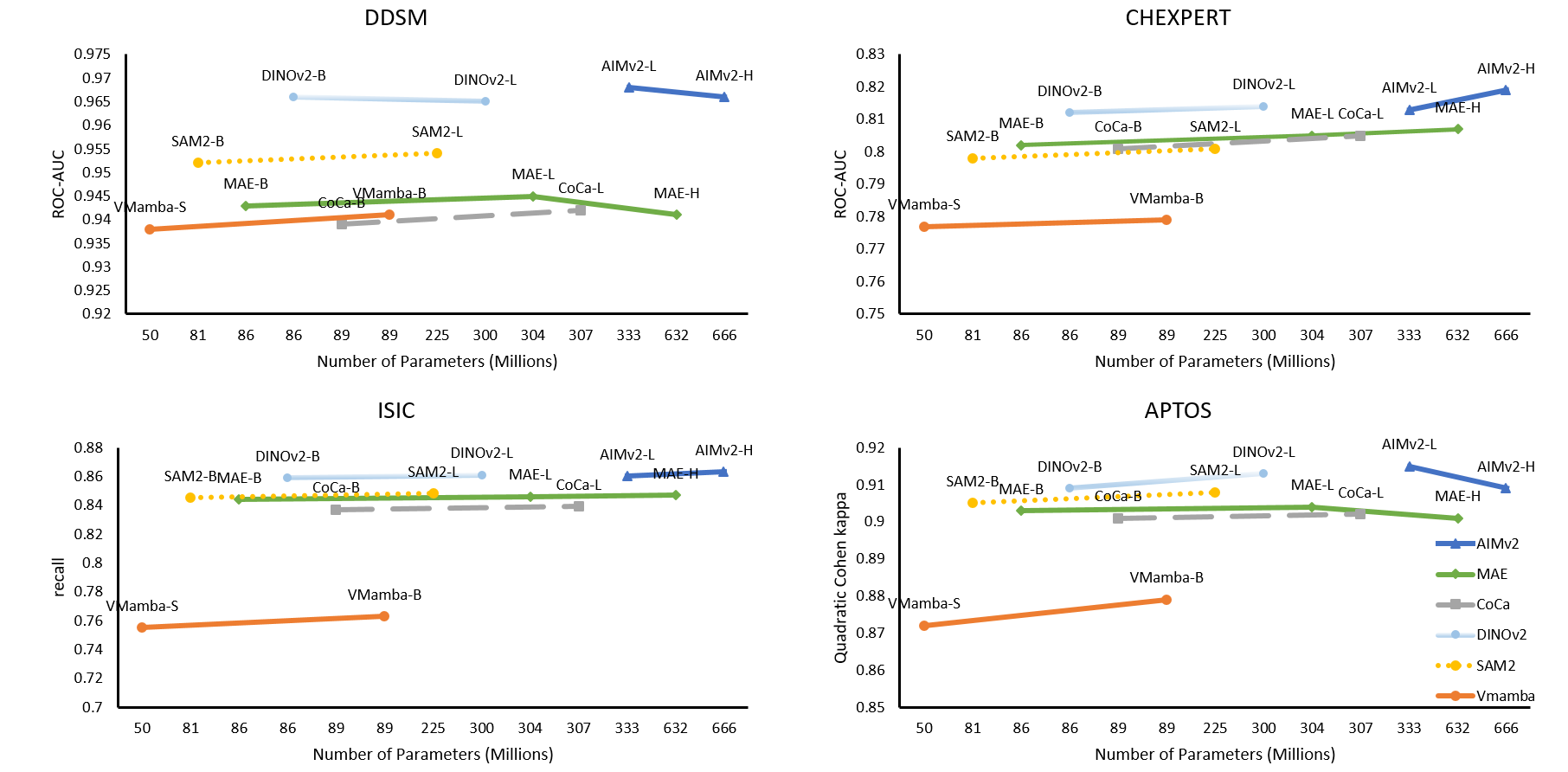}
\caption{Comparison of different sizes of backbones in each foundation model across all datasets, based on the linear head of the unfreeze configuration.}
\label{fig1}
\end{figure*}

\subsection{Datasets and Evaluation metrics}
To evaluate the performance of foundation models across diverse medical domains, we selected four distinct medical datasets, each belonging to different organs. These datasets were chosen to conduct a comprehensive assessment of the models' capabilities in handling a variety of medical classification tasks. Furthermore, each dataset employs unique evaluation metrics suited to its specific characteristics and clinical applications, ensuring a thorough and representative analysis of the models' performance.

•	\textbf{CBIS-DDSM}\cite{lee2017curated}: The CBIS-DDSM dataset is an updated and standardized version of the Digital Database for Screening Mammography (DDSM)\cite{Heath2001DDSM}. It comprises a curated subset of 2,620 scanned film mammography studies, including normal, benign, and malignant cases with verified pathology information. The dataset is widely used for developing and testing decision support systems in mammography. The AUC-ROC (Area Under the Receiver Operating Characteristic Curve) metric, which measures a model’s ability to distinguish between different classes with higher values indicating better discrimination, is used to evaluate this dataset.

•	\textbf{ISIC2019}\cite{codella2018skin}: The ISIC2019 dataset is a comprehensive and diverse collection of dermoscopic images, specifically curated for skin lesion classification tasks. It includes 25,331 images, spanning nine diagnostic categories: melanoma, melanocytic nevus, basal cell carcinoma, actinic keratosis, benign keratosis, dermatofibroma, vascular lesion, squamous cell carcinoma, and seborrheic keratosis. This dataset provides a robust platform for developing and evaluating algorithms aimed at accurately diagnosing various skin conditions. Recall is used as the evaluation metric for this dataset, which measures the model's ability to correctly identify positive cases (e.g., correctly classifying malignant lesions), ensuring that critical cases are not missed in clinical settings.

•	\textbf{APTOS2019}\cite{karthik2019aptos}: The APTOS2019 dataset is a comprehensive collection of retinal images used for diabetic retinopathy detection and grading. It includes 3,660 training images and additional images for validation and testing, with ratings based on the International Clinical Diabetic Retinopathy severity scale (ICDR). The Quadratic Cohen Kappa is the chosen metric for assessing this dataset, which measures the agreement between the predicted and actual ratings of diabetic retinopathy severity.

•	\textbf{CHEXPERT}\cite{irvin2019chexpert}: The CHEXPERT dataset is one of the largest and most comprehensive collections of chest radiographs, designed to facilitate advancements in automated chest X-ray interpretation. It includes 224,316 chest radiographs from 65,240 patients, featuring uncertainty labels and radiologists-labeled reference standards. The dataset covers a wide range of pathologies, including cardiomegaly, edema, consolidation, atelectasis, and pleural effusion. For this dataset, the evaluation metric is ROC-AUC.

\section{EXPERIMENTS}
In this section, we detail the comprehensive experimental setup and results of the introduced models and medical image datasets discussed in the previous section.
\subsection{Implementation Details}
\label{Implementation}
As mentioned earlier, two distinct configurations were employed for designing the classification head. We used both linear and multi-layer attention heads to convert the models' final features into classification scores. For the attention head, we used a pre-trained transformer [12], and as input, we provided the patch embeddings of the final layers of the foundation models. For linear classification, we used [CLS] embeddings. However, for models like SAM2 without the [CLS] token, we utilized the average pooling of patch features to obtain the [CLS] embedding. Additionally, to better analyze the models' capabilities, we evaluated the results in two scenarios: 1) with the model backbone frozen and only the classification head trained, and 2) with the entire model fine-tuned.

For DDSM and CheXpert, the image channels were set to 1 since they are grayscale images. To handle this, instead of increasing the image channels, similar to \cite{huix2024natural}, we modified the patch embeddings of the models to work with single-channel images. Since all models were pre-trained with RGB 3-channel images, we removed the weights of the second and third channels and used only the weights of the first channel for the model's patch embeddings.

For a fair comparison, the input image sizes were set to $224 \times 224$ for all models. For models trained with different input sizes, we used bicubic interpolation to adjust their position embeddings \cite{dosovitskiy2020image}. Standard augmentations were applied to the training images, including normalization, vertical and horizontal flips, color jitter, and random resize crops. Each data set was divided into training sets (80\%), validation sets (10\%), and test sets (10\%). In order to fine-tune the models, supervised learning was employed with the AdamW optimizer. To determine the optimal learning rate, we searched in a range of $10^{-3}$ to $10^{-5}$.

\subsection{Model Performance Comparison} As shown in Tables \cref{ddsm,aptos,isic,chexpert}, In our evaluation of various foundation models for fine-tuning medical tasks, AIMv2 achieved the highest performance across both frozen and unfrozen configurations in all four datasets. Following AIMv2, the DINOv2 and SAM2 models also delivered excellent results.

An essential consideration is the size of the datasets. For larger datasets like CHEXPERT and ISIC2019, the models faced fewer challenges during training. For smaller datasets such as CBIS-DDSM and APTOS2019, the models were still able to achieve high performance, demonstrating their robustness and adaptability.

\subsection{Linear and Multi Attention Head} During frozen fine-tuning, where only the linear or attention head mechanisms were fine-tuned, the multi-layer attention mechanism achieved superior results across all four datasets. To further assess the adaptability of foundation models to medical tasks, we fine-tuned the models by unfreezing their weights, utilizing both a linear head and a multi-layer attention head as detailed in Section \ref{Implementation}. In the unfrozen configuration, the linear head achieved superior results compared to its performance in the frozen mode, often matching or exceeding the performance of multi-attention layers across all models.

\subsection{Model Size} As part of our approach, we also experimented with different variations of the backbone to determine the best version of each model. As demonstrated in \cref{fig1}, \cref{ddsm,aptos}, for smaller datasets such as APTOS2019 and CBIS-DDSM, the base and large versions of the model achieved the highest performance. Conversely, for larger datasets like ISIC2019 and CHEXPERT, as indicated in Tables \cref{isic} and \ref{chexpert}, models with a larger number of parameters exhibited superior results.

\subsection{Training Time} We assessed the training times of various models. Training time is defined as the number of iterations required for the validation metrics to achieve peak performance, normalized across datasets. Training a linear head on a frozen foundation significantly extended the duration compared to other configurations. In contrast, using a multi-attention head in a frozen mode yielded the shortest training time among all configurations.

While fine-tuning the whole model, both linear and attention heads exhibited similar training times. Notably, the CoCa model had the longest training time. On the other hand, the AIMv2, DINOv2, and SAM2 models not only delivered better performance but also required less training time. Among these, the MAE and SAM2 models stood out, demonstrating the lowest overall training time.

\section{Conclusion}
This study highlights the significant impact of state-of-the-art foundation models, including DINOv2, MAE, VMamba, CoCa, SAM2, AIMv2, and CLIP, on medical image classification. By fine-tuning these models on diverse datasets such as CBIS-DDSM, ISIC2019, APTOS2019, and CHEXPERT, we have demonstrated their capability to improve classification performance. Our experiments revealed that using a multi-layer attention head consistently outperformed linear heads during frozen fine-tuning, while unfreezing the models led to substantial performance gains for both linear and multi-layer attention heads. Among the models, AIMv2 showed the highest performance across all datasets, followed closely by DINOv2 and SAM2. These findings underscore the robustness and adaptability of foundation models in handling various medical imaging tasks, highlighting their potential to improve diagnostic accuracy and enhance overall clinical practice. Furthermore, our results indicate that advancements in natural domain training positively influence medical image classification, suggesting that integrating cutting-edge foundation models can drive meaningful progress in the medical field.

\bibliographystyle{IEEEtran}
\bibliography{refs}

\begin{thebibliography}{10}
\providecommand{\url}[1]{#1}
\csname url@samestyle\endcsname
\providecommand{\newblock}{\relax}
\providecommand{\bibinfo}[2]{#2}
\providecommand{\BIBentrySTDinterwordspacing}{\spaceskip=0pt\relax}
\providecommand{\BIBentryALTinterwordstretchfactor}{4}
\providecommand{\BIBentryALTinterwordspacing}{\spaceskip=\fontdimen2\font plus
\BIBentryALTinterwordstretchfactor\fontdimen3\font minus
  \fontdimen4\font\relax}
\providecommand{\BIBforeignlanguage}[2]{{%
\expandafter\ifx\csname l@#1\endcsname\relax
\typeout{** WARNING: IEEEtran.bst: No hyphenation pattern has been}%
\typeout{** loaded for the language `#1'. Using the pattern for}%
\typeout{** the default language instead.}%
\else
\language=\csname l@#1\endcsname
\fi
#2}}
\providecommand{\BIBdecl}{\relax}
\BIBdecl

\bibitem{bommasani2021opportunities}
R.~Bommasani, D.~A. Hudson, E.~Adeli, R.~Altman, S.~Arora, S.~von Arx, M.~S.
  Bernstein, J.~Bohg, A.~Bosselut, E.~Brunskill \emph{et~al.}, ``On the
  opportunities and risks of foundation models,'' \emph{arXiv preprint
  arXiv:2108.07258}, 2021.

\bibitem{radford2019language}
A.~Radford, J.~Wu, R.~Child, D.~Luan, D.~Amodei, I.~Sutskever \emph{et~al.},
  ``Language models are unsupervised multitask learners,'' \emph{OpenAI blog},
  vol.~1, no.~8, p.~9, 2019.

\bibitem{kenton2019bert}
J.~D. M.-W.~C. Kenton and L.~K. Toutanova, ``Bert: Pre-training of deep
  bidirectional transformers for language understanding,'' in \emph{Proceedings
  of naacL-HLT}, vol.~1, no.~2.\hskip 1em plus 0.5em minus 0.4em\relax
  Minneapolis, Minnesota, 2019.

\bibitem{radford2021learning}
A.~Radford, J.~W. Kim, C.~Hallacy, A.~Ramesh, G.~Goh, S.~Agarwal, G.~Sastry,
  A.~Askell, P.~Mishkin, J.~Clark \emph{et~al.}, ``Learning transferable visual
  models from natural language supervision,'' in \emph{International conference
  on machine learning}.\hskip 1em plus 0.5em minus 0.4em\relax PMLR, 2021, pp.
  8748--8763.

\bibitem{he2022masked}
K.~He, X.~Chen, S.~Xie, Y.~Li, P.~Doll{\'a}r, and R.~Girshick, ``Masked
  autoencoders are scalable vision learners,'' in \emph{Proceedings of the
  IEEE/CVF conference on computer vision and pattern recognition}, 2022, pp.
  16\,000--16\,009.

\bibitem{oquab2023dinov2}
M.~Oquab, T.~Darcet, T.~Moutakanni, H.~Vo, M.~Szafraniec, V.~Khalidov,
  P.~Fernandez, D.~Haziza, F.~Massa, A.~El-Nouby \emph{et~al.}, ``Dinov2:
  Learning robust visual features without supervision,'' \emph{arXiv preprint
  arXiv:2304.07193}, 2023.

\bibitem{liu2024vmamba}
Y.~Liu, Y.~Tian, Y.~Zhao, H.~Yu, L.~Xie, Y.~Wang, Q.~Ye, J.~Jiao, and Y.~Liu,
  ``Vmamba: Visual state space model,'' \emph{arXiv preprint arXiv:2401.10166},
  2024.

\bibitem{ravi2024sam}
N.~Ravi, V.~Gabeur, Y.-T. Hu, R.~Hu, C.~Ryali, T.~Ma, H.~Khedr, R.~R{\"a}dle,
  C.~Rolland, L.~Gustafson \emph{et~al.}, ``Sam 2: Segment anything in images
  and videos,'' \emph{arXiv preprint arXiv:2408.00714}, 2024.

\bibitem{fini2024multimodal}
E.~Fini, M.~Shukor, X.~Li, P.~Dufter, M.~Klein, D.~Haldimann, S.~Aitharaju,
  V.~G.~T. da~Costa, L.~B{\'e}thune, Z.~Gan \emph{et~al.}, ``Multimodal
  autoregressive pre-training of large vision encoders,'' \emph{arXiv preprint
  arXiv:2411.14402}, 2024.

\bibitem{yu2022coca}
J.~Yu, Z.~Wang, V.~Vasudevan, L.~Yeung, M.~Seyedhosseini, and Y.~Wu, ``Coca:
  Contrastive captioners are image-text foundation models,'' \emph{arXiv
  preprint arXiv:2205.01917}, 2022.

\bibitem{matsoukas2023pretrained}
C.~Matsoukas, J.~F. Haslum, M.~S{\"o}derberg, and K.~Smith, ``Pretrained vits
  yield versatile representations for medical images,'' \emph{arXiv preprint
  arXiv:2303.07034}, 2023.

\bibitem{matsoukas2022makes}
C.~Matsoukas, J.~F. Haslum, M.~Sorkhei, M.~S{\"o}derberg, and K.~Smith, ``What
  makes transfer learning work for medical images: Feature reuse \& other
  factors,'' in \emph{Proceedings of the IEEE/CVF Conference on Computer Vision
  and Pattern Recognition}, 2022, pp. 9225--9234.

\bibitem{huix2024natural}
J.~P. Huix, A.~R. Ganeshan, J.~F. Haslum, M.~S{\"o}derberg, C.~Matsoukas, and
  K.~Smith, ``Are natural domain foundation models useful for medical image
  classification?'' in \emph{Proceedings of the IEEE/CVF Winter Conference on
  Applications of Computer Vision}, 2024, pp. 7634--7643.

\bibitem{shi2023generalist}
P.~Shi, J.~Qiu, S.~M.~D. Abaxi, H.~Wei, F.~P.-W. Lo, and W.~Yuan, ``Generalist
  vision foundation models for medical imaging: A case study of segment
  anything model on zero-shot medical segmentation,'' \emph{Diagnostics},
  vol.~13, no.~11, p. 1947, 2023.

\bibitem{wang2022medclip}
Z.~Wang, Z.~Wu, D.~Agarwal, and J.~Sun, ``Medclip: Contrastive learning from
  unpaired medical images and text,'' \emph{arXiv preprint arXiv:2210.10163},
  2022.

\bibitem{mansoori2024polyp}
M.~Mansoori, S.~Shahabodini, J.~Abouei, K.~N. Plataniotis, and A.~Mohammadi,
  ``Polyp sam 2: Advancing zero shot polyp segmentation in colorectal cancer
  detection,'' \emph{arXiv preprint arXiv:2408.05892}, 2024.

\bibitem{yi2023towards}
H.~Yi, Z.~Qin, Q.~Lao, W.~Xu, Z.~Jiang, D.~Wang, S.~Zhang, and K.~Li, ``Towards
  general purpose medical ai: Continual learning medical foundation model,''
  \emph{arXiv preprint arXiv:2303.06580}, 2023.

\bibitem{mansoori2024self}
M.~Mansoori, S.~Shahabodini, J.~Abouei, K.~N. Plataniotis, and A.~Mohammadi,
  ``Self-prompting polyp segmentation in colonoscopy using hybrid yolo-sam 2
  model,'' \emph{arXiv preprint arXiv:2409.09484}, 2024.

\bibitem{SAM2-MIS}
Y.~Zhang and Z.~Shen, ``Unleashing the potential of sam2 for biomedical images
  and videos: A survey,'' \emph{arXiv preprint arXiv:2408.12889}, 2024.

\bibitem{dosovitskiy2020image}
A.~Dosovitskiy, ``An image is worth 16x16 words: Transformers for image
  recognition at scale,'' \emph{arXiv preprint arXiv:2010.11929}, 2020.

\bibitem{lin2014microsoft}
T.-Y. Lin, M.~Maire, S.~Belongie, J.~Hays, P.~Perona, D.~Ramanan,
  P.~Doll{\'a}r, and C.~L. Zitnick, ``Microsoft coco: Common objects in
  context,'' in \emph{Computer Vision--ECCV 2014: 13th European Conference,
  Zurich, Switzerland, September 6-12, 2014, Proceedings, Part V 13}.\hskip 1em
  plus 0.5em minus 0.4em\relax Springer, 2014, pp. 740--755.

\bibitem{lee2017curated}
R.~S. Lee, F.~Gimenez, A.~Hoogi, K.~K. Miyake, M.~Gorovoy, and D.~L. Rubin, ``A
  curated mammography data set for use in computer-aided detection and
  diagnosis research,'' \emph{Scientific data}, vol.~4, no.~1, pp. 1--9, 2017.

\bibitem{karthik2019aptos}
M.~Karthik and S.~Dane, ``Aptos 2019 blindness detection,'' \emph{Kaggle
  https://kaggle. com/competitions/aptos2019-blindness-detection Go to
  reference in}, p.~5, 2019.

\bibitem{deng2009imagenet}
J.~Deng, W.~Dong, R.~Socher, L.-J. Li, K.~Li, and L.~Fei-Fei, ``Imagenet: A
  large-scale hierarchical image database,'' in \emph{2009 IEEE conference on
  computer vision and pattern recognition}.\hskip 1em plus 0.5em minus
  0.4em\relax Ieee, 2009, pp. 248--255.

\bibitem{gu2023mamba}
A.~Gu and T.~Dao, ``Mamba: Linear-time sequence modeling with selective state
  spaces,'' \emph{arXiv preprint arXiv:2312.00752}, 2023.

\bibitem{codella2018skin}
N.~C. Codella, D.~Gutman, M.~E. Celebi, B.~Helba, M.~A. Marchetti, S.~W. Dusza,
  A.~Kalloo, K.~Liopyris, N.~Mishra, H.~Kittler \emph{et~al.}, ``Skin lesion
  analysis toward melanoma detection: A challenge at the 2017 international
  symposium on biomedical imaging (isbi), hosted by the international skin
  imaging collaboration (isic),'' in \emph{2018 IEEE 15th international
  symposium on biomedical imaging (ISBI 2018)}.\hskip 1em plus 0.5em minus
  0.4em\relax IEEE, 2018, pp. 168--172.

\bibitem{irvin2019chexpert}
J.~Irvin, P.~Rajpurkar, M.~Ko, Y.~Yu, S.~Ciurea-Ilcus, C.~Chute, H.~Marklund,
  B.~Haghgoo, R.~Ball, K.~Shpanskaya \emph{et~al.}, ``Chexpert: A large chest
  radiograph dataset with uncertainty labels and expert comparison,'' in
  \emph{Proceedings of the AAAI conference on artificial intelligence},
  vol.~33, no.~01, 2019, pp. 590--597.

\bibitem{Heath2001DDSM}
M.~Heath, K.~Bowyer, D.~Kopans, R.~Moore, and W.~P. Kegelmeyer, ``The digital
  database for screening mammography,'' in \emph{Proceedings of the Fifth
  International Workshop on Digital Mammography}, 2001, pp. 212--218.

\end{thebibliography}

\end{document}